# High Rectification Ratio at Room Temperature in Rhenium(I) Compound

Subas Rajbangshi[ab#], Nila Pal[a#], Robinur Rahman[b], Vladimir N. Nesterov[c], Lisa Roy[d], Shishir Ghosh[b,*], Prakash Chandra Mondal[a*]

[a]*Department of Chemistry, Indian Institute of Technology, Kanpur, Uttar Pradesh 208016, India*

[b]*Department of Chemistry, Jahangirnagar University, Dhaka 1342, Bangladesh*

[c]*Department of Chemistry, University of North Texas, 1155 Union Circle, Box 305070, TX 76203, USA*

[d]*Institute of Chemical Technology Mumbai-IOC Odisha Campus Bhubaneswar, Bhubaneswar 751013, India*

[#]These two authors equally contributed to this work.

E-mail: shishir.ghosh@juniv.edu (SG); pcmondal@iitk.ac.in (PCM)

**Abstract**

Electrical current rectification is an interesting electronic feature, popularly known as a diode. Achieving a high rectification ratio in a molecular junction has been a long-standing goal in molecular electronics. The present work describes mimicking electrical current rectification with π-stacked rhenium(I) compound sandwiched between two electrical contacts. Among the two mononuclear rhenium compounds studied here, $[Re(CO)_4(PPh_3)\{\kappa^1\text{-(N)-saccharinate}\}]$ (**1**) and $[Re(CO)_3(\kappa^2\text{-phen})\{\kappa^1\text{-(N)-saccharinate}\}]$ (**2**), the latter show strong π–π interactions-induced high rectification ratio of ~ $4\times10^3$ at ±2.0 V at room temperature. Alternating current (AC)-based electrical measurements ensuring AC to DC electrical signal conversion at a frequency *f* of 1 KHz showing **2** can act as an excellent half-wave rectifier. Asymmetric charge injection barrier height at the electrode/Re(I) interfaces of the devices with a stacking configuration of $p^{++}$-Si/Re compound$_{31nm}$(**2**)/ITO originates the flow of electrical current unidirectionally. The charge transport mechanism governed by thermally activated hopping phenomena, and charge carrier propagation is explained through an energy profile considering the Fermi levels of two electrodes, and the energy of frontier molecular orbitals, HOMO, and LUMO, confirming rectification is of a molecular origin. The present work paves the way to combine different organometallic compounds as circuit elements in nanoelectronic devices to achieve numerous exciting electronic features.

**Introduction**

A theoretical framework modeled by Aviram and Ratner in which donor and acceptor organic moieties are separated by a saturated (σ bridge) group, can orginiate electrical rectification that gives rise birth to 'molecular electronics'.[1] Molecular electronics aim at understanding charge transport phenomena in molecular junctions (MJs) where either a single molecule or self-assembled monolayers or oligomer films are sandwiched between two conductors.[2–6] The MJs are fabricated by varying molecular functionalities, structures, compositions, orientations, deposition methods, and various top and bottom electrode configurations followed by understanding the origin of exciting electronic functions.[7–10] Molecular electronics lie at interfaces of chemistry, physics, materials, and engineering where size reduction of charge transport media along with the controlling conductance at the molecular level is the major goal.[11–15] Molecular electronics enable us to understand the effect of humidity, temperature, solvent, light, and magnetic field in the MJs as the electrical response may drastically change in response to those external stimuli.[16–22] The most important aspect of molecular electronics is to understand the charge transport mechanism with the varied thickness of molecular layers that separate the two electrical contacts, which might be influenced by the temperature.[11,23–25] Among the electronic functions that a molecular junction may yield, rectification is considered one of the promising features that allows electrical currents to flow in one direction but block the current in the reverse bias polarity. Such behavior are observed with various organic molecules-based electronic devices.[2,26–30] The reason of such asymmetric current-voltage response includes asymmetric nature at the electrode-molecules interfaces, or different frontier molecular orbitals.[31–35] Such rectification with either coordination compounds or organometallic compounds is rarely explored inspite of their (i) capacity of both the metal atoms or ions and the ligands to produce active sites for charge transport through redox activity, (ii) tunablity of the frontier molecular orbitals and energy gaps, (iii) possibility of various molecular structures based on the geometry around the metal centres, (iv) varied electrical dipole-moment and directions, (v) metal-ligand back-bonding between the d to π* orbitals, and (vi) enhanced stability.[36–39] Many of MJs are fabricated using metallic bottom contacts such as gold, or silver, which are expensive and not many devices are utilized in AC to DC signal conversion, which is actually required for practical utilization. The present work utilizes Re(I)-based organometallic compounds as circuit elements in the MJs that originates high electrical current rectification. There are several advantages with Re(I)-based compounds including forming neutral, two, and three dimensional metallacycles to supramolecular architectures that reveal exciting photophysical, and redox-activities.[40] The presence of strong π–π interactions between sac-sac and phen-phen in **2**, leads to high electrical current rectification, while **1** having weaker π-π interactions shows symmetric current at both polarity. Our present study creates an opportunity to study organometallic compounds in molecular junctions that could produce many promising electronic functions.

**Results and discussion**

To understand the molecular signature in an asymmetric current-voltage response, we consider two different Re(I)-based compounds, one with less (**1**) and another with more (**2**) intermolecular π-π stacking. A labile mononuclear rhenium complex [Re(CO)$_4$(NCMe){κ$^1$-(N)-sac}] was reacted with PPh$_3$ and 1,10-phenanthroline (phen) to afford [Re(CO)$_4$(PPh$_3$){κ$^1$-(N)-sac}] (**1**) and [Re(CO)$_3$(κ$^2$-phen){κ$^1$-(N)-sac}] (**2**) in 51 and 77% yields, respectively (**Scheme S1**). The crystal structures of **1** (CCDC No. 2236790) and **2** (CCDC No. 2236791) are depicted in **Figure 1a,c**, and their corresponding unit cell and packing are shown in **Figure 1b,d**, respectively (**Figure S1a,b, Table S1**). **1** contains Re(I) ligated by four carbonyls (CO), a triphenylphosphine (PPh$_3$), and a saccharinate (sac) ligand. **2** consists of Re(I) coordinated by a sac, a phen, and three carbonyl ligands. In both cases, CO are bound to Re(I) in a terminal fashion, and the three CO in **2** adopt a facial arrangement around the metal center occupying one face of the octahedron. The sac ligand is coordinated to rhenium(I) via the nitrogen in a κ$^1$-fashion laying *cis* to the PPh$_3$ ligand in **1** and *cis* to phen ligand in **2**. The Re(1)–P(1) [2.5028(6) Å] and Re(1)–N(1) [2.183(2) Å] bond distances in **1** are similar to those reported data.[41,42] For **2,** the Re–N bond distances involving both the sac [Re(1)–N(1) 2.205(3) Å and Re(2)–N(4) 2.214(3) Å, for two independent molecules] and phen [Re(1)–N(2) 2.170(3) Å and Re(1)–N(3) 2.182(3) Å; Re(2)–N(5) 2.175(3) Å and Re(2)–N(6) 2.182(3) Å]. The evidence of structural distortion is found from the P(1)–Re(1)–N(1) of 88.79(6)° for **1** and the small bite-angle of the bidentate phen ligand [N(2)–Re(1)–N(3) 76.05(10)°] for **2**. This observation is also supported by hybrid DFT studies at B3LYP/SDD(Re),6-31G**(rest) (see **S6-7**). The P(1)-Re(1)-N(1) angle in a monomer of **1** is calculated to be 88.83° which is in well-agreement to the crystal structure. Similarly, the calculated N(2)–Re(1)–N(3) bite angle in **2** is 74.89°, predicting the observed distortion in the octahedral arrangement. The symmetry-independent molecules in noncentrosymmetric structures with Z′ = 2 often interact via motifs that are 'inversion favouring'.[43] The solid-state structure reveals strong π–π interactions between sac-sac and phen-phen in **2** (C-C distances are 3.368 – 3.508 Å), while it is weaker in **1** (C-C distances are 3.400 – 3.535 Å). Such π–π interactions lower HOMO-LUMO energy gap which eventually could greatly affect photophysical properties including many striking electronic features.[44] Using the same optimized geometry, the theoretically obtained HOMO of **1** ~ -6.23 eV, **2** ~ -5.73 eV, and LUMO of **1** ~ -1.63 eV, **2** ~ -2.60 eV and the corresponding HOMO-LUMO gaps are 4.60 and 3.13 eV, respectively (**Figure 2a,d**). It can be observed that the HOMO is primarily metal centered in both the compounds, while the LUMO is ligand-centered (**Figure 2, and S2**), indicating the possibility of metal-to-ligand charge transfer. The molecular electrostatic potential surface (MEPS) analysis reveals that the positive potential zone is centered around oxygen atoms bonded with sulphur and carbon atoms of the sac ligand, rest of the constituents of the molecule are contributing to negative potential for **1**. While, the oxygen atoms of carbonyl ligands are also contributing to the positive potential zones along with the oxygen atoms of sac-ligand of **2**, thus give rise to a higher electric dipole moment of 10.21 D vs. 7.39 D for **2** compared to **1** (**Figure 2e-g**).

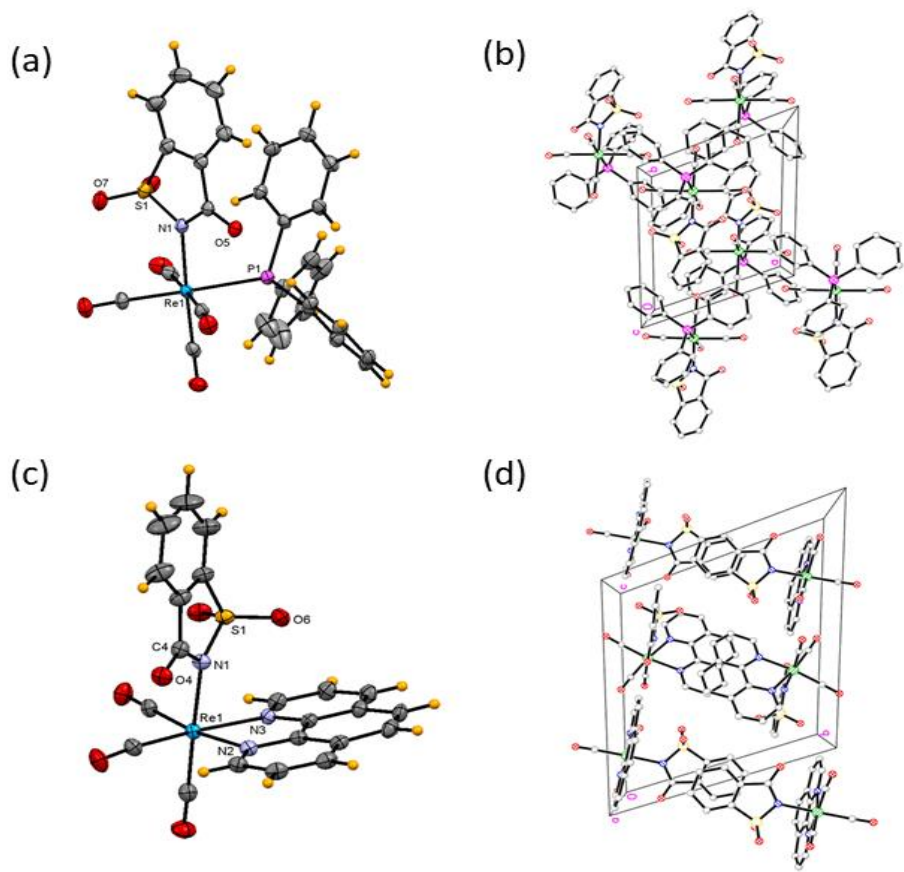

**Figure 1**. (a), (c) General view of [Re(CO)$_4$(PPh$_3$){$\kappa^1$-(N)-sac}] (**1**) and [Re(CO)$_3$($\kappa^2$-phen){$\kappa^1$-(N)-sac}] (**2**) showing 50% probability atomic displacement ellipsoids, with selected bond distances (Å) and angles (°): (a) Re(1)–P(1) 2.5028(6), Re(1)–N(1) 2.183(2), P(1)–Re(1)–N(1) 88.79(6); (c) Re(1)–N(1) 2.205(3), Re(1)–N(2) 2.170(3), Re(1)–N(3) 2.182(3), N(1)–Re(1)–N(2) 81.38(10), N(1)–Re(1)–N(3) 84.77(10), N(2)–Re(1)–N(3) 76.05(10). Fragments of the packing diagrams of **1** and **2** are shown in Fig. (b), and (d), respectively.

The $^1$H NMR spectrum of both Re(I) complexes exhibits a series of multiplets in the aromatic region, attributed to the sac and phenyl/pyridyl protons (**Figure S3, S4**), while the $^{31}$P{$^1$H} NMR spectrum of **1** displays a singlet at δ 9.9 ppm that signifies the presence of phosphine (**Figure S5**). The FT-IR spectrum of both compounds shows vibrational stretching frequencies at 1674 cm$^{-1}$ for **1** and 1680 cm$^{-1}$ for **2**, attributed to CO in sac ligand, while the rhenium-bound CO appears between 2110-1915 cm$^{-1}$ for **1** and 2027-1910 cm$^{-1}$ for **2** (**Figure S6**). DFT calculations predict that the Re-CO distances in **1** are 1.95, 1.98, 2.02 and 2.03 Å. Similarly, the Re-CO distances in **2** are calculated at 1.93, 1.94 and 1.95 Å for the three facial CO ligands. Moreover, the C-O bond distances are slightly increased in **1** (1.14 - 1.16 Å) and **2** (1.16 Å), as compared to an unbound CO (1.137 Å). Our theoretical calculations also predict that the CO ligands are perpendicular to each other, with C-Re-C bond angles in the range of 88.7 – 90.4° in **1** and 89.9 – 90.1° in **2**. Re(I)-bound CO groups are predicted to have vibrational frequencies in the range of 2095-2028 cm$^{-1}$ for **1** and 2101-2007 cm$^{-1}$ for **2**. A shift in lower CO stretching frequency in Re(I) compound than that of free CO ($v$ 2143 cm$^{-1}$) suggest that the carbonyl groups are influenced by the combined simultaneous effect of the d$\pi$ electron density of Re(I), and the $\pi$-back bonding (to $\pi$* orbitals of CO).[45,46] The CO stretching frequencies of **2** are shifted toward lower energy with respect to those of **1**, since Re(I) of the former is electron rich as compared to the latter; thus, stronger π-back-donation from Re(I) to CO group occurs in **2** which weakens the CO stretching. The electrospray mass spectrum of **1** contains a major peak corresponding to [M+Na]$^+$ at m/z 766, and for **2**, the isotope envelopes correspond to [M+H]$^+$ at m/z 634 (**Figure S7a,b**). Thermogravimetric analysis shows weight loss curve of crystal samples at 307°C for **1** and 390°C for **2** (**Figure S8**), respectively, indicating high-temperature device processability and operational.

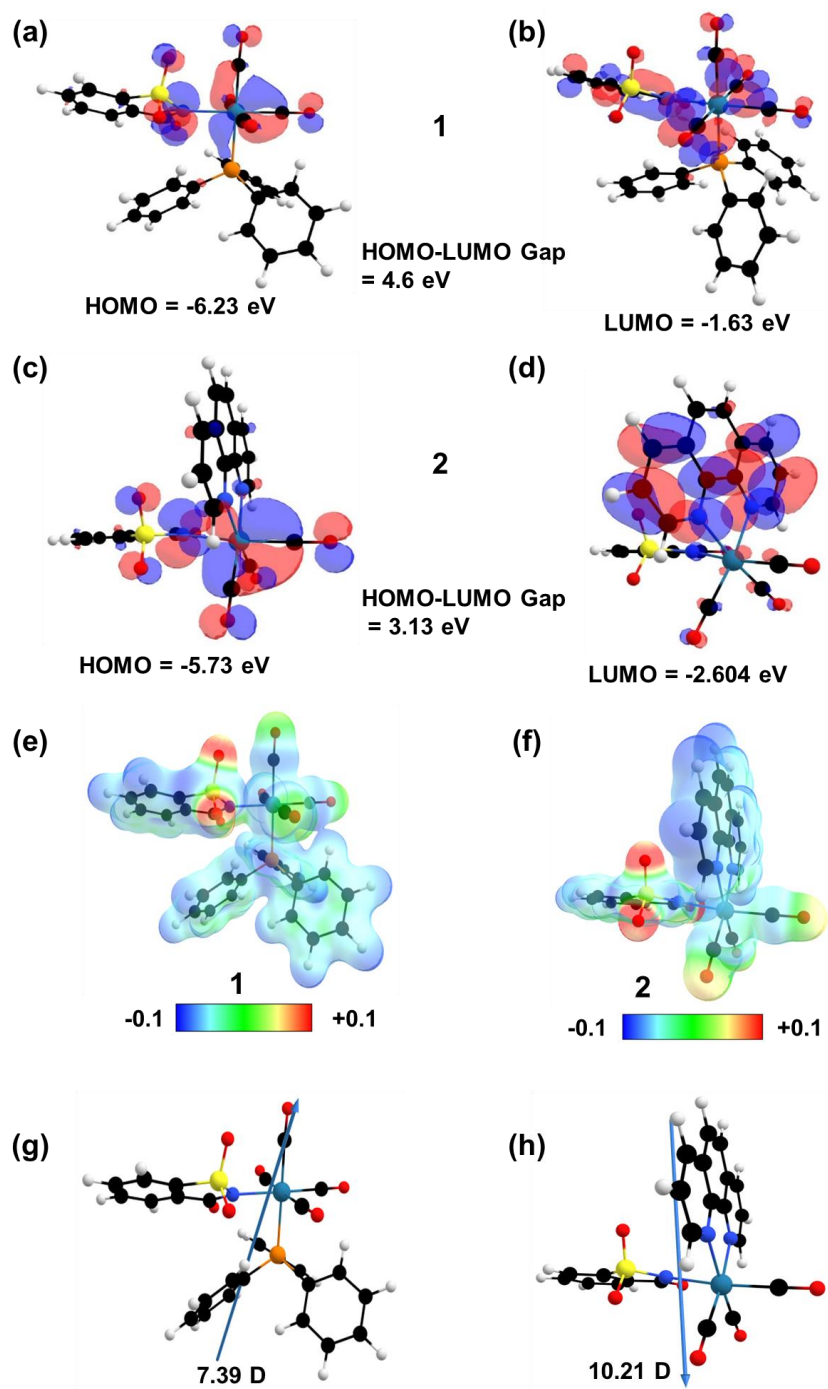

**Figure 2**. DFT-derived Frontier orbital plots of **1** and **2** using B3LYP/(SDD (Re), 6-31G** (rest) level of theory. (a) HOMO, (b) LOMO positioned at -6.23 eV and -1.63 eV, respectively for **1**, (c) HOMO, (d) LOMO positioned at -5.73 eV and -2.64 eV, respectively for **2**. Molecular Electrostatic Potential (MESP) with contour value = 0.08. Blue (strongly negative); Green (weakly negative); Red (strongly positive). The MESP plots of both compounds (e) **1**, and (f) **2**. Molecular dipole moments of (g) 7.39 D for **1**, (h) 10.21 D for **2**.

The electronic absorption spectrum of **1** and **2** shows one strong band at λ$_{max}$ = 268 nm with ε = 1.4 × 10$^3$ M$^{-1}$cm$^{-1}$ and at λ$_{max}$ = 274 nm with ε = 5.2 × 10$^3$ M$^{-1}$cm$^{-1}$, respectively (**Figure 3a,b**) that are attributed to the ligand-based π → π* electronic transitions. Several weak absorptions are also observed at 276, 298, and 322 nm for **1** and 257, 294 (sh), and 392 nm for **2**, respectively, (**Table S2**). The lowest energy transitions appear at λ$_{max}$ = 322 nm with ε = 0.78 × 10$^3$ M$^{-1}$cm$^{-1}$ for **1** and λ$_{max}$ = 392 nm with ε = 0.74 × 10$^3$ M$^{-1}$cm$^{-1}$ for **2**, ascribed to MLCT (metal to ligand charge transfer) electronic transition, d$_π$(Re) → π*$_L$. Similar absorption spectra were reported with different Re(I) compounds earlier.[47] The solid-state UV-vis spectra of thin films of **1** and **2** prepared on freshly cleaned quartz substrates are recorded for deducing λ$_{onset}$ values appear at 337.5 nm and 525 nm for **1** and **2**, respectively (**Figure S9a,b**). The estimated HOMO-LUMO energy gaps are 3.67 and 2.36 eV for **1** and **2**, which are slightly different from the gas-phase theoretical calculations. Upon UV-irradiation, **2** shows marked luminescence with λ$_{em}$ = 580 nm (excitation λ$_{ex}$ at 392 nm, **Figure S10**), the emissive state is typically $^3$MLCT (dπ(Re) → π*(N^N)), and due to the lack of such emissive state of **1** is found to be non-luminescent. To understand the electrochemical charge transfer dynamics, and kinetic parameters, electrochemical measurements were carried out in a static conditions, where electron transfer between the working electrode, and the Re(I) compounds (concentration of 1 mM) at the solid-liquid interfaces was recorded in a polar aprotic medium. A potentiostat was used to modulate the energy of the Fermi level of a working electrode (glassy carbon electrode, area 0.07 cm$^2$), whose potential was controlled by a reference electrode (Ag/AgNO$_3$), while the output current was measured between working electrode and a platinum wire as a counter electrode. **1** shows two successive one-electron reduction occurring at −1.41 and −1.78 V vs. Ag/AgNO$_3$ attributed to the ligand-based reduction process forming [Re(CO)$_4$(PPh$_3$){κ$^1$-(N)-sac}]$^{•−}$ which is a well-known redox phenomenon.[48] The corresponding oxidation appears at −1.10, and −1.53 V vs. Ag/AgNO$_3$ on the return scan (**Figure 2c**). However, **1** shows a metal ion-based redox process at −1.57 V vs. Ag/AgNO$_3$, which can be attributed to the high energy barrier (E$_F$-HOMO) between the working electrode Fermi level (E$_F$) and HOMO of **1** (**Table S3**). **2** undergoes a chemically irreversible oxidation process due to one electron transferred at +1.08 V vs. Ag/AgNO$_3$ assigned to d$^6$ to d$^5$ transition, i.e, Re(I) to Re(II) oxidation (**Figure 2d**). At a more anodic potential such as +1.6 V, an intense oxidation signal could arise from an electrochemical-chemical-electrochemical (ECE) process forming dimer by coupling two neutral radicals [Re(CO)$_3$($\kappa^2$-phen)]$^{\bullet}$ generated from [Re(CO)$_3$($\kappa^2$-phen){$\kappa^1$-(N)-sac}] after first oxidation.[49,50] However, this mechanism remains debatable, as others consider that there could be an alternative path to the intramolecular fast ligand substitution reaction. Akin to other Re(I)-tricarbonyl complexes bearing a polypyridyl ligand, **2** also shows two reduction waves at –1.27 V and –1.78 V vs. Ag/AgNO$_3$.[62] The first reduction is ligand-based, with the added electron accommodated on 1,10-phen, leading to the formation of radical anion [Re(CO)$_3$($\kappa^2$-phen){$\kappa^1$-(N)-sac}]$^{\bullet-}$ which is a quasireversible redox process.[62] The second reduction leads to the dissociation of sac ligand to generate [Re(CO)$_3$($\kappa^2$-phen)]$^{-}$.[62] The cyclic voltammograms of **1** and **2** with varying scan rates (10 to 500 mV s$^{-1}$, **Figure S11a,** and **Figure S12a**). The peak current densities of these two compounds increase linearly with the square root of the scan rates (R$^2$ = 0.99, **Figure S11b, Figure S12b**), quite common for freely diffusing electroactive analytes; thus no surface attachment was noticed even after a longer CV run. The heterogeneous electron transfer rate constants, and diffusion coefficient (we consider the first ligand redox couple) for **1**, and **2** are estimated at 47.38 s$^{-1}$ (for **1**), 77.94 s$^{-1}$ (for **2**), and 1.27×10$^{-4}$ cm$^2$s$^{-1}$ (anodic for **1**), 1.45×10$^{-3}$ cm$^2$s$^{-1}$ (cathodic for **1**), and 3.66× 10$^{-6}$ cm$^2$s$^{-1}$ (anodic for **2**), 7.03×10$^{-6}$ cm$^2$s$^{-1}$ (cathodic for **2**) (**Table S4**). Larger peak-to-peak separation ($\Delta E_p$ > 57 mV at 25°C and increases with scan rates) between cathodic and anodic processes ensure the redox processes are electrochemically quasi-reversible, meaning slow electrode kinetics due to the slower electron transfer at the working electrode/analyte interfaces than the mass transport. Here, mass transport is considered only diffusion controlled; neither convection (as the cell is unstirred), nor migration (as the concentration of the supporting electrolyte is much higher than the analytes to balance the charge at the electrode surface) is involved, thus the overall process is sluggish and follows a non-Nernstian electrochemical process.

Thin films of **1** and **2** are prepared on freshly cleaned p$^{++}$-Si substrates (native oxides were removed using hydrofluoric acid) via spin-coating (concentration of 10$^{-5}$ M in CH$_3$CN, spinning speed 4000 rpm for 40 s) followed by drying before recording the XPS, and device fabrication (see **S15**). The thickness of the films was measured via optical profilometry and estimated at ~ 31±2 nm (**Figure S13**). The field-emission scanning electron microscopy images of films **1** and **2** show uniformly distributed circular disc morphology with average diameters of 0.29 ± 0.03 µm and 0.4 ± 0.05 µm and average interparticle distances of 0.91 ± 0.30 µm and 0.95 ± 0.23 µm, respectively (**Figure S14a-b**). The survey scan and high-resolution XPS signals are included in supporting information (**Figure S15-S17**). The binding energies for Re 4f$_{7/2}$ are 41.2 and 41.1 eV; for 4f$_{5/2}$ are 43.53 and 43.38 eV for **1** and **2**, respectively (**Figure S17a, b**), which shows good agreement with the previous reports.[51,52] Both spectra consist of two deconvoluted peaks with positions near 401.8 eV assigned for C-N, N-(C=O) bonds, and 399 eV assigned for organic matrix environment and bonding with Re(I) metal ion.[52,53] As Re (1.9 in pouling scale) has very similar electronegativity as Co (1.88 in pouling scale), so Re-N binding energy contribution in N 1s spectra can be considered at the same position. The smaller curve area at binding energy ~399 eV can be understand by the high electronegativity of N (3.04 in pouling scale) with compared to Re metal.

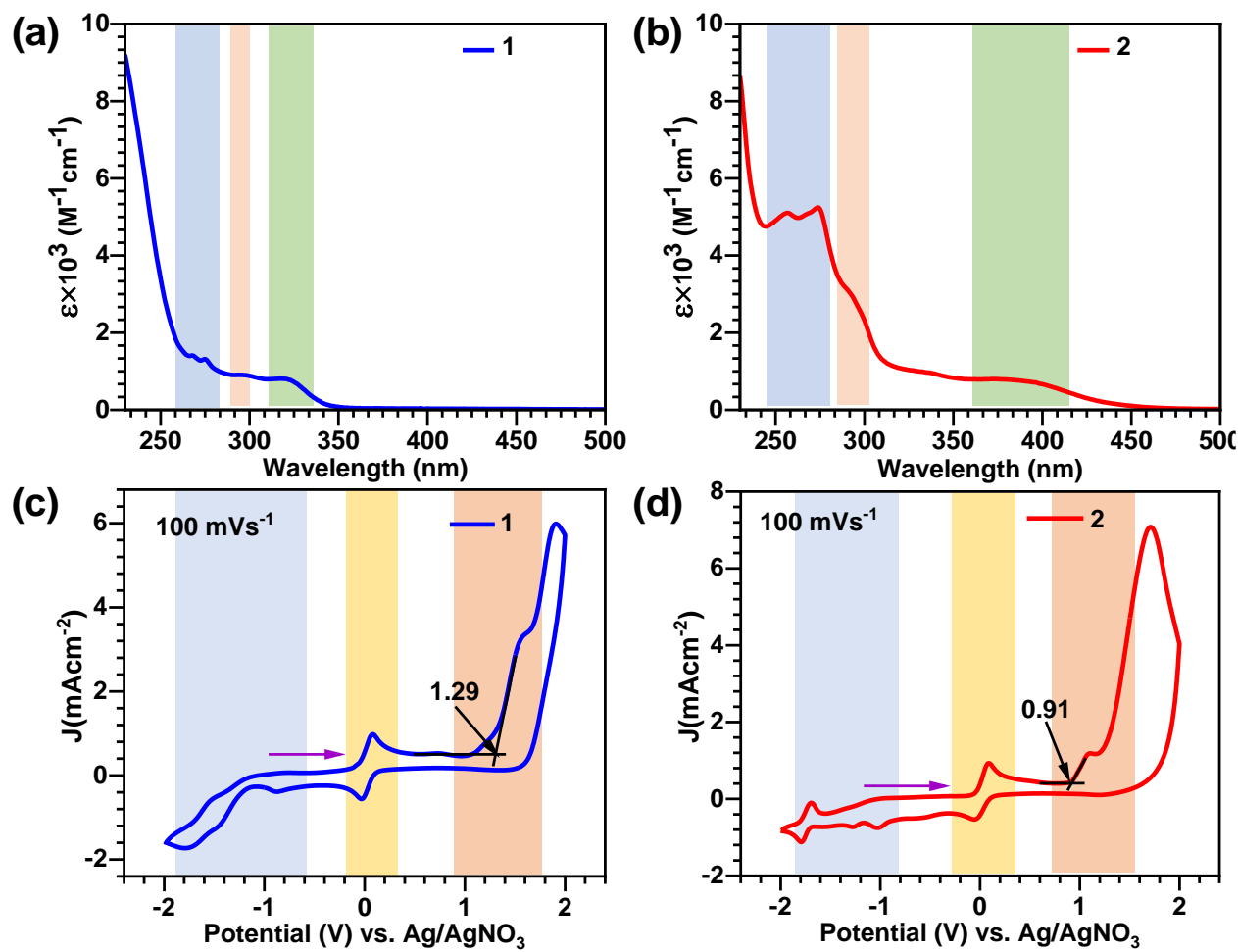

**Figure 3**. (a)-(b) UV-vis absorption spectrum of **1** (blue) and **2** (red) recorded in CH$_3$CN at room temperature at a concentration of $1 \times 10^{-5}$ M, (c) to (d) cyclic voltammograms of **1** (blue), and **2** (red) (1 mM of concentration of **1** and **2** in dry MeCN with 0.1 M supporting Bu$_4$NPF$_6$ electrolyte) at 100 mVs$^{-1}$. A solution of ferrocene (Fc) was added to the electrochemical cell as an internal standard.

The schematic of p$^{++}$-Si/**1** or **2**/ITO MJs fabrication flow chart and the image of actual device are shown in **Figure 4a,b**. Highly boron-doped Si substrates were used as bottom electrodes (Orientation <100>, NANOSHEL, resistivity ~0-100 Ω.cm, thickness of 430 µm), and the top contact was ITO (dimension 1.4 cm × 0.5 cm, thickness ~ 100 ± 5 nm) **1** or **2** was sandwiched between the electrodes. An external voltage was applied at the top ITO, with respect to the bottom p$^{++}$-Si . The voltage was swept from −1.0 V to 1.0 V and varied up to ±5.0 V. From the current vs. voltage plots (I-V) of device **2** in linear and semi-logarithmic scales, more efficient charge transport in the negative bias region was noticed, which resembles p-type rectifier behavior with a lower threshold voltage of − 0.4 V (in single sweeps, **Figure 4c,d**). The electrical current rectification ratio, RR, defined as equation (i) is used for the quantification of asymmetric current response at specific bias,

$$RR = \left|\frac{I(V_-)}{I(V_+)}\right| \qquad (i)$$

A plot of RR vs. applied voltage is shown in **Figure 4e,** and the mean rectification ratio of the device was obtained 4.3(±0.5)×10$^3$ at ±2.0 V. To ensure device integrity, a statistic was made with 22 devices fabricated in different batches (batch 1-3) and more than 70% yield was achieved (**Table S5-S6**). The I-V sweeps of device **1** shows a symmetric response at the both single and double sweep with negligible RR ~1 at ±2.0 V (**Figure S18a-c**). The double sweeps of the linear, semi-logarithmic curves, and statistical conductance values for both devices are shown in **Figure S19** and **20a-c**. The device **1** shows higher conductance (~267.88 ± 24.78 µS) than device **2** (~79.64 ± 3.29 µS). When a voltage is applied in the bottom p$^{++}$-Si with respect to the top ITO electrode device **2** shows opposite behavior to that of the earlier one, while device **1** shows a symmetric response (**Figure S21a-b**).

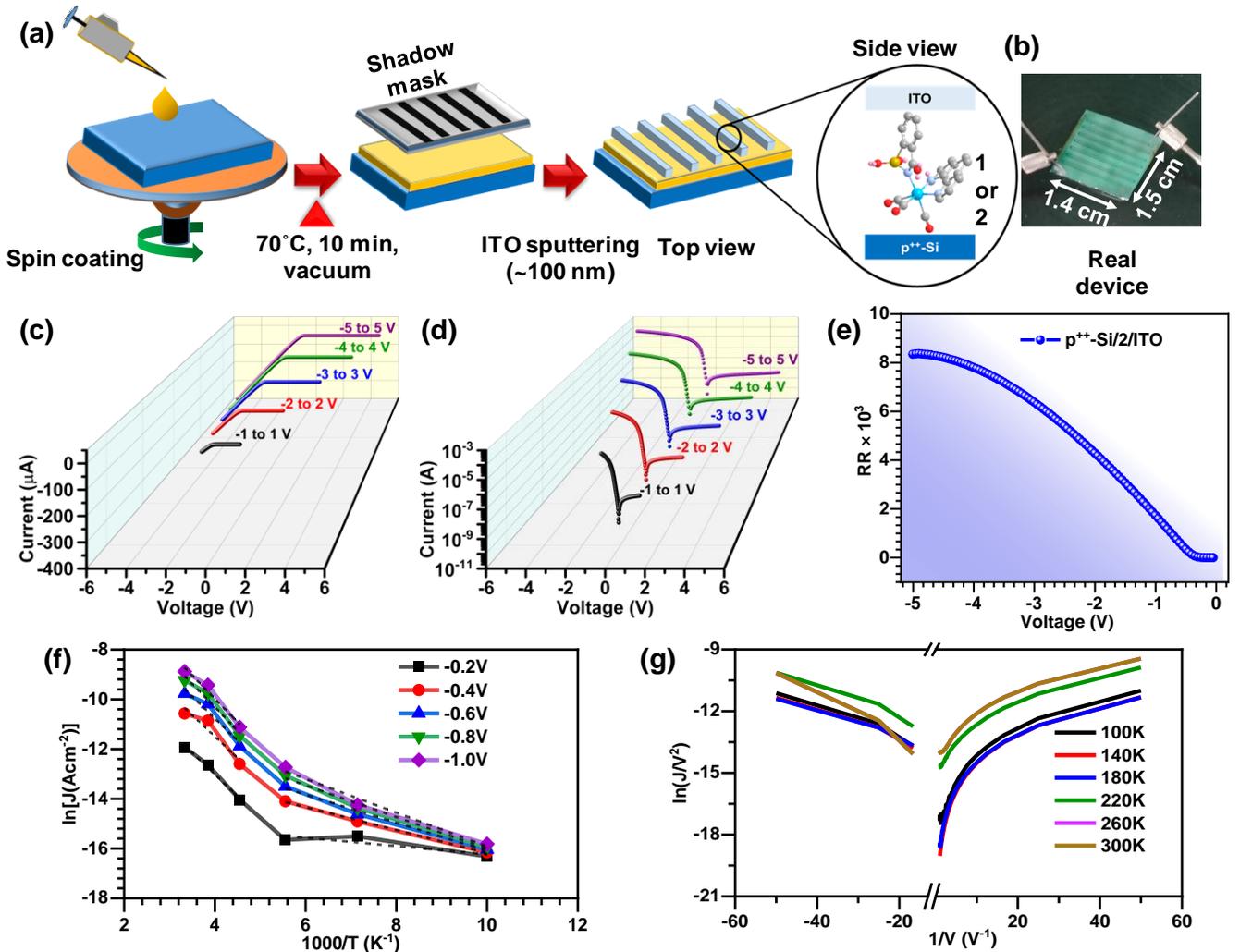

**Figure 4** (a) Schematic diagram of two terminal molecular junctions made with either **1** or **2**, the zoomed view shows the cross-section of p$^{++}$-Si/**1** or **2**/ITO, (b) the real device image under probed condition, (c) current-voltage (I-V) characteristics of device **2** in linear scale, (d) semi-logarithmic scale with a single sweep, (e) variation of rectification ratio with an applied voltage of **2**, (f) Arrhenius plots, and (g) ln (J/V$^2$) vs 1/V curve for device **2** for different temperature at ±1 V.

An ohmic response from the reference devices (without incorporating molecules) confirms that the rectifier nature of device **2** has molecular origin but is not because of the Schottky barrier formation between two electrodes (**Figure S21c-d**). The activation energy ($E_a$) required for charges to propagate in the device was obtained from the Arrhenius plots at varied temperature ranges (100 K to 300 K), as shown in **Figure 4f** for device **2** (**Table S8**). The Arrhenius plots reveal two distinguishable temperature ranges: lower one is in the temperature range of 100-180 K, and the higher one from 200 to 300 K. The value of $E_a$ becomes lower at a low operating voltage in low-temperature regions (~ 58 meV at –1 V, and ~14 meV at –0.2 V). The higher $E_a$ at higher-temperature regions is attributed to the thermally activated hopping transport mechanism, which is well-established for molecular electronics.[54] Further, the non-linear nature of the $\ln(J/V^2)$ vs. $1/V$ curves suggests that the charge transport is not dominated by Fowler-Nordheim (FN) tunneling (**Figure 4g**). To check the rectification property of the devices with an AC ($V_{input} = V_A \sin(2\pi ft)$, $V_A$ is the amplitude) signal as input, and output was measured across 180 kΩ series resistance (circuit design is depicted in **Figure 5a,b**). The applied voltage amplitude ($V_A$) is kept sufficiently high to supply the devices with an input DC voltage, $V_{DC} \approx 0.8$ V, and the series resistance is kept in the circuit to prevent electrical breakdown of the devices. A similar configuration was used by others.[55] Output profiles of device **2** ensure the conversion of AC to DC transformation either in a positive or negative direction, according to the application of input signal to the respective electrodes from a frequency range of 50 Hz to 1 kHz, at 10 kHz input signal almost passes through the device (**Fig 5c,d**). Moreover, the direction of output signals matches well with the DC measurement. However, the output DC voltages were found below the 3 dB voltage throughout the frequency spectrum (**Figure S22**). The AC measurement with device **1**, was found to pass all the input signals irrespective of its application either at bottom p$^{++}$-Si or top ITO electrode (**Figure S23**). A theoretical cut-off frequency can be predicted by the relation,

$$f = \frac{1}{2\pi RC} \qquad (ii)$$

where, R is resistance applied in series and C is the capacitance of the molecular films. An electrical impedance spectrum is essential to deduce the individual electrical parameters of the MJs.[56–58] From the impedance spectroscopy analysis of device **2**, the areal capacitance of the molecular films obtained at 37.1 pF.cm$^{-2}$ (**Figure S24, Table S7**). Taking the series resistance value of 180 kΩ and the capacitance value of 2.2 pF, the theoretical cutoff frequency is calculated at 3.4×10$^5$ Hz. To depict the energy profile of the devices, experimentally determined HOMO, and LUMO values are used. The Fermi level of the heavily boron-doped Si and conducting ITO was considered at –5.0 eV ($E_{F1}$) and –4.3 eV ($E_{F2}$), respectively.[59–61] The bottom electrodes are in contact with the molecules with stronger van der Waals forces compared to the top elecrodes, so stronger coupling between the molecules and the bottom electrodes is expected. The energy band diagram depicts that though HOMO of **2** lies in between the electrodes and it has energy barriers of ~ 0.7 eV with $E_{F1}$ of $p^{++}$-Si, and ~ 1.4 eV with $E_{F2}$ of ITO at isolated conditions (**Figure 6a**). At contact position (at V= 0 V) the whole system has an aligned Fermi level, maintaining the potential equilibrium (**Figure 6b**). Now considerg the bottom $p^{++}$-Si electrode as the reference, when -2 V (this voltage is preferred to design the model as we have achieved a mean rectification ratio at ± 2V) is applied to the device, both electrodes will experience half of the applied bias according to the potentiometric division rule (**Figure 6c**).[62] A nonlinear potential profile (dotted sky color) is assumed here considering the theoretically higher dipole moments of the materials. Due to the strong dipolar interaction, the bottom electrode and HOMO will be closer together. So, at -2 V, the carrier conduction is promoted through HOMO because of its easy accessibility. Conversely, at +2 V, the ITO electrode is pulled up and $p^{++}$-

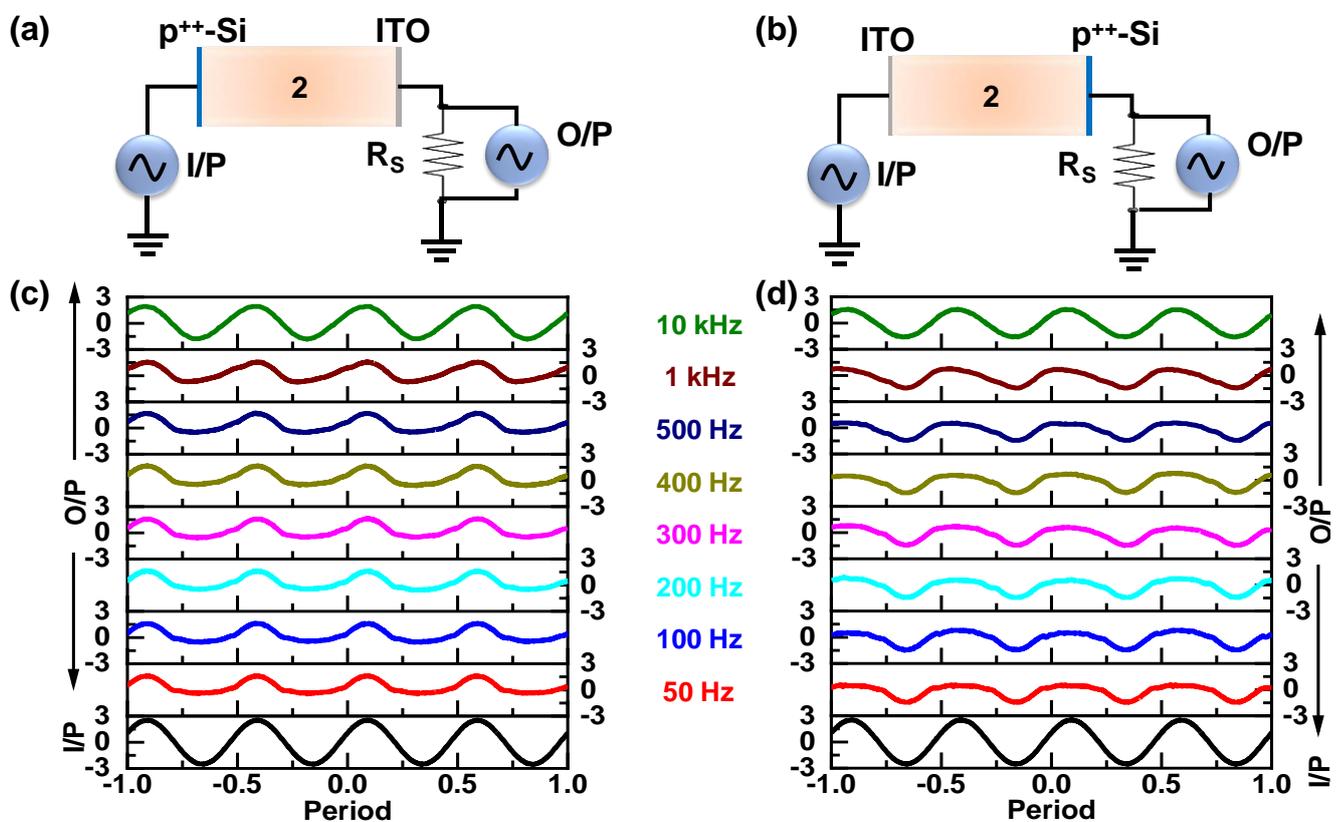

**Figure 5**. (a) and (b) Experimental set to measure the rectifier's frequency (*f*) response and output curves (output was measured across of 180 kΩ load) at *f* from 50 Hz to 10 kHz, when input is given to (c) bottom $p^{++}$-Si, and (d) top ITO electrode.

Si goes down (**Figure 6d**). Even so, the HOMO follows the potential profile, but still, a considerably high potential barrier to shifted $E_{F2}$ restricts carrier flow through the HOMO. In the same way, when the ITO electrode is kept as a reference, high carrier flow is facilitated at +2 V, and impeded at -2 V (**Figure S25**). This variation of current in ± 2 V applied bias gives rise to a higher RR for **2**. But in the case of device **1**, the HOMO of **1** (- 6.06 eV) lies in the vicinity of $E_{F2}$, so carrier flow is enabled at both bias polarities regardless of which electrode is kept as a reference (**Figure S26-S27**). A symmetric I-V response was obtained from the I-V sweeps.

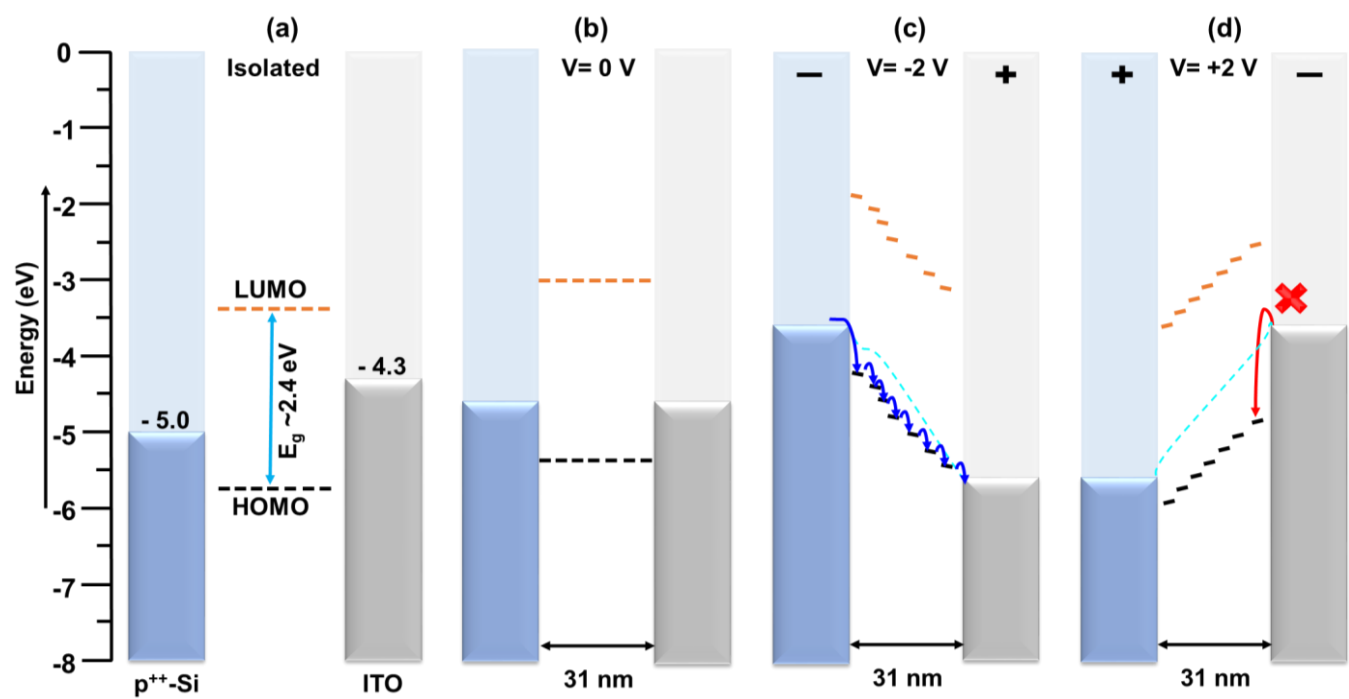

**Figure 6**. Schematic of the charge transport mechanism of device **2** (bottom p$^{++}$-Si electrode is the reference one) (a) Isolated, (b) 0 V, (c) +2 V, and (c) -2 V.

## Conclusion

One of the major goals of molecular electronics is to achieve high RR (>10$^3$) at low bias and to be able to operate at high-frequency (>1 kHz) AC to DC signal conversion. Towards this goal, we successfully incorporate Re(I)-based organometallic compounds having different electric dipole moments, and π-π stacking. Such types of compounds are hardly utilized in molecular electronics; thus, we focus on an unexplored molecular system for understanding charge transport and mimicking an attractive electronic feature such as a diode. The rectification of Re(I)-based devices was tested as a half-wave rectifier by amplifying AC inputs at varied frequencies besides the asymmetric DC-based electrical measurements. Such observation unequivocally confirms working efficiency at high frequency (> 1 KHz). Such features are encouraging, as commercial diodes work in higher frequency ranges. To support the transport mechanism, temperature-dependent charge conduction at different operational bias has been the key charge carrier propagation. The molecular diodes can be further improved in the context of their stability, and even higher RR values.

## ASSOCIATED CONTENT

**Supporting Information**

Experimental procedures; FT-IR data; $^1$H, $^{31}$P{$^1$H} NMR; XPS; crystal data; and structure refinement; TGA data; solid-state UV-Vis data; CV plots; additional I-V plots; AC measurements; electrical impedance data; and computational details (PDF).

## AUTHOR INFORMATION

**SR:** experimental, data analysis, main draft writing, **NP:** experimental, data analysis, main draft writing, **RR:** synthesis, **VNN:** XRD measurement, **LR:** DFT study, **SG:** draft writing, supervision, **PCM:** main draft writing, and editing, supervision, funding.

**Notes**

The authors declare no competing financial interest.

## ACKNOWLEDGEMENT

SR acknowledges the Department of Science and Technology, Indian National Science Academy (DST INSA), Government of India for the prestigious Indian Research Fellowship (ISRF, Grant No. INSA/CHM/2022427) to short term research stay at IIT Kanpur. NP is thankful to IIT Kanpur for IPDF fellowship (PDF-313). PCM acknowledges financial support from the Council of Scientific & Industrial Research, project NO.:01(3049)/21/EMR-II, New Delhi, India. SG and SR acknowledge the Ministry of Education, Government of the People's Republic of Bangladesh. The authors acknowledge IIT Kanpur for infrastructures and equipment facilities. Authors are thankful to Dr. A. Vilan and Prof. C. Frisbie for their critical comments that help enrich the quality of the draft.

## REFERENCES

[1] A. Aviram, M. A. Ratner, *Chem. Phys. Lett.* **1974**, *29*, 277.
[2] R. Gupta, J. A. Fereiro, A. Bayat, A. Pritam, M. Zharnikov, P. C. Mondal, *Nat. Rev. Chem.* **2023**, *7*, 106.
[3] D. A. Egger, F. Rissner, E. Zojer, G. Heimel, *Adv. Mater.* **2012**, *24*, 4403.
[4] T. Wang, X. Zhang, Z. Wang, X. Zhu, J. Liu, X. Min, T. Cao, X. Fan, *Polymers (Basel).* **2019**, *11*, 1564.
[5] L. Li, C. R. Prindle, W. Shi, C. Nuckolls, L. Venkataraman, *J. Am. Chem. Soc.* **2023**, *145*, 18182.
[6] M. W. Gu, C. T. Lai, I. C. Ni, C. I. Wu, C. hsien Chen, *Angew. Chemie - Int. Ed.* **2023**, *62*, DOI 10.1002/anie.202214963.
[7] Y. Zhu, Y. Zhou, L. Ren, J. Ye, H. Wang, X. Liu, R. Huang, H. Liu, J. Liu, J. Shi, P. Gao, W. Hong, *Angew. Chemie* **2023**, *135*, DOI 10.1002/ange.202302693.
[8] Q. Pi, D. Bi, D. Qiu, H. Wang, X. Cheng, Y.-Q. Feng, Q. Zhao, M. Zhou, *J. Mater. Chem. C* **2021**, DOI 10.1039/D1TC01425K.
[9] Z. Wang, H. Dong, T. Li, R. Hviid, Y. Zou, Z. Wei, X. Fu, E. Wang, Y. Zhen, K. Nørgaard, B. W. Laursen, W. Hu, *Nat. Commun.* **2015**, *6*, DOI 10.1038/ncomms8478.
[10] G. D. Kong, J. Jang, S. Choi, G. Lim, I. S. Kim, T. Ohto, S. Maeda, H. Tada, H. J. Yoon, *Small* **2023**, DOI 10.1002/smll.202305997.
[11] L. Venkataraman, J. E. Klare, C. Nuckolls, M. S. Hybertsen, M. L. Steigerwald, *Nature*

[12] B. Capozzi, J. Xia, O. Adak, E. J. Dell, Z. F. Liu, J. C. Taylor, J. B. Neaton, L. M. Campos, L. Venkataraman, *Nat. Nanotechnol.* **2015**, *10*, 522.
[13] J. Tang, Y. Wang, J. E. Klare, G. S. Tulevski, S. J. Wind, C. Nuckolls, *Angew. Chemie Int. Ed.* **2007**, *46*, 3892.
[14] G. Koplovitz, G. Leitus, S. Ghosh, B. P. Bloom, S. Yochelis, D. Rotem, F. Vischio, M. Striccoli, E. Fanizza, R. Naaman, D. H. Waldeck, D. Porath, Y. Paltiel, *Small* **2019**, *15*, 1804557.
[15] P. C. Mondal, N. Kantor-Uriel, S. P. Mathew, F. Tassinari, C. Fontanesi, R. Naaman, *Adv. Mater.* **2015**, *27*, 1924.
[16] P. Chandra Mondal, U. M. Tefashe, R. L. McCreery, *J. Am. Chem. Soc.* **2018**, *140*, 7239.
[17] S. H. Choi, C. Risko, M. Carmen Ruiz Delgado, B. Kim, J. L. Brédas, C. Daniel Frisbie, *J. Am. Chem. Soc.* **2010**, *132*, 4358.
[18] A. Vilan, D. Cahen, *Chem. Rev.* **2017**, *117*, 4624.
[19] X. Yao, X. Sun, F. Lafolet, J.-C. Lacroix, *Nano Lett.* **2020**, *20*, 6899.
[20] H. Atesci, V. Kaliginedi, J. A. Celis Gil, H. Ozawa, J. M. Thijssen, P. Broekmann, M. A. Haga, S. J. Van Der Molen, *Nat. Nanotechnol.* **2018**, *13*, 117.
[21] P. C. Mondal, P. Roy, D. Kim, E. E. Fullerton, H. Cohen, R. Naaman, *Nano Lett.* **2016**, *16*, 2806.
[22] R. Torres-Cavanillas, G. Escorcia-Ariza, I. Brotons-Alcázar, R. Sanchís-Gual, P. C. Mondal, L. E. Rosaleny, S. Giménez-Santamarina, M. Sessolo, M. Galbiati, S. Tatay, A. Gaita-Ariño, A. Forment-Aliaga, S. Cardona-Serra, *J. Am. Chem. Soc.* **2019**, *142*, 17572.
[23] A. Vilan, D. Aswal, D. Cahen, *Chem. Rev.* **2017**, *117*, 4248.
[24] D. Rambabu, A. E. Lakraychi, J. Wang, L. Sieuw, D. Gupta, P. Apostol, G. Chanteux, T. Goossens, K. Robeyns, A. Vlad, *J. Am. Chem. Soc.* **2021**, jacs. 1c04591.
[25] M. Kulke, D. M. Olson, J. Huang, D. M. Kramer, J. V. Vermaas, *Small* **2023**, DOI 10.1002/smll.202304013.
[26] G. D. Kong, S. E. Byeon, J. Jang, J. W. Kim, H. J. Yoon, *J. Am. Chem. Soc.* **2022**, *144*, 7966.
[27] P. Song, S. Guerin, S. J. R. Tan, H. V. Annadata, X. Yu, M. Scully, Y. M. Han, M. Roemer, K. P. Loh, D. Thompson, C. A. Nijhuis, *Adv. Mater.* **2018**, *30*, 1706322.
[28] D. D. James, A. Bayat, S. R. Smith, J.-C. Lacroix, R. L. McCreery, *Nanoscale Horiz.* **2017**, *3*, 45.
[29] C. Van Dyck, M. A. Ratner, *Nano Lett.* **2015**, *15*, 1577.
[30] K. S. Wimbush, W. F. Reus, W. G. van der Wiel, D. N. Reinhoudt, G. M. Whitesides, C. A. Nijhuis, A. H. Velders, *Angew. Chemie Int. Ed.* **2010**, *49*, 10176.
[31] J. Shin, S. Yang, Y. Jang, J. S. Eo, T. W. Kim, T. Lee, C. H. Lee, G. Wang, *Nat. Commun.* **2020**, *11*, 1412.
[32] C. N. Verani, *Dalt. Trans.* **2018**, *47*, 14153.
[33] A. Z. Thong, M. S. P. Shaffer, A. P. Horsfield, *Sci. Rep.* **2018**, *8*, 9120.
[34] W. Peng, N. Chen, C. Wang, Y. Xie, S. Qiu, S. Li, L. Zhang, Y. Li, *Angew. Chemie Int. Ed.* **2023**, *62*, DOI 10.1002/anie.202307733.
[35] H. Jeong, D. Kim, D. Xiang, T. Lee, *ACS Nano* **2017**, *11*, 6511.
[36] J. G. Park, B. A. Collins, L. E. Darago, T. Runčevski, M. E. Ziebel, M. L. Aubrey, H. Z. H. Jiang, E. Velasquez, M. A. Green, J. D. Goodpaster, J. R. Long, *Nat. Chem.* **2021**, *13*, 594.
[37] U. Rashid, E. Chatir, L. Medrano Sandonas, P. Sreelakshmi, A. Dianat, R. Gutierrez, G. Cuniberti, S. Cobo, V. Kaliginedi, *Angew. Chemie* **2023**, *135*, DOI 10.1002/ange.202218767.
[38] P. Sachan, P. C. Mondal, *Analyst* **2020**, *145*, 1563.
[39] P. Sachan, P. Chandra Mondal, *ChemElectroChem* **2020**, *7*, 4186.
[40] D. Gupta, P. Rajakannu, B. Shankar, R. Shanmugam, F. Hussain, B. Sarkar, M. Sathiyendiran, *Dalt. Trans.* **2011**, *40*, 5433.

[41] M. R. Moni, S. Ghosh, S. M. Mobin, D. A. Tocher, G. Hogarth, M. G. Richmond, S. E. Kabir, *J. Organomet. Chem.* **2018**, *871*, 167.
[42] S. E. Kabir, F. Ahmed, S. Ghosh, M. R. Hassan, M. S. Islam, A. Sharmin, D. A. Tocher, D. T. Haworth, S. V. Lindeman, T. A. Siddiquee, D. W. Bennett, K. I. Hardcastle, *J. Organomet. Chem.* **2008**, *693*, 2657.
[43] R. Taylor, J. C. Cole, C. R. Groom, *Cryst. Growth Des.* **2016**, *16*, 2988.
[44] M. Dharmarwardana, B. M. Otten, M. M. Ghimire, B. S. Arimilli, C. M. Williams, S. Boateng, Z. Lu, G. T. McCandless, J. J. Gassensmith, M. A. Omary, *Proc. Natl. Acad. Sci. U. S. A.* **2021**, *118*, DOI 10.1073/pnas.2106572118.
[45] Z. Nour, H. Petitjean, D. Berthomieu, *J. Phys. Chem. C* **2010**, *114*, 17802.
[46] M. Frank, L. Jürgensen, J. Leduc, D. Stadler, D. Graf, I. Gessner, F. Zajusch, T. Fischer, M.-A. Rose, D. N. Mueller, S. Mathur, *Inorg. Chem.* **2019**, *58*, 10408.
[47] B. Shankar, S. Sahu, N. Deibel, D. Schweinfurth, B. Sarkar, P. Elumalai, D. Gupta, F. Hussain, G. Krishnamoorthy, M. Sathiyendiran, *Inorg. Chem.* **2014**, *53*, 922.
[48] C. J. Stanton, C. W. Machan, J. E. Vandezande, T. Jin, G. F. Majetich, H. F. Schaefer, C. P. Kubiak, G. Li, J. Agarwal, *Inorg. Chem.* **2016**, *55*, 3136.
[49] A. Carreño, E. Solís-Céspedes, C. Zúñiga, J. Nevermann, M. M. Rivera-Zaldívar, M. Gacitúa, A. Ramírez-Osorio, D. Páez-Hernández, R. Arratia-Pérez, J. A. Fuentes, *Chem. Phys. Lett.* **2019**, *715*, 231.
[50] F. Paolucci, M. Marcaccio, C. Paradisi, S. Roffia, C. A. Bignozzi, C. Amatore, *J. Phys. Chem. B* **1998**, *102*, 4759.
[51] H. Hori, K. Koike, M. Ishizuka, K. Takeuchi, T. Ibusuki, O. Ishitani, *J. Organomet. Chem.* **1997**, *530*, 169.
[52] E. M. Johnson, R. Haiges, S. C. Marinescu, *ACS Appl. Mater. Interfaces* **2018**, *10*, 37919.
[53] K. Artyushkova, B. Kiefer, B. Halevi, A. Knop-Gericke, R. Schlogl, P. Atanassov, *Chem. Commun.* **2013**, *49*, 2539.
[54] L. Luo, L. Balhorn, B. Vlaisavljevich, D. Ma, L. Gagliardi, C. D. Frisbie, *J. Phys. Chem. C* **2014**, *118*, 26485.
[55] J. Trasobares, D. Vuillaume, D. Théron, N. Clément, *Nat. Commun.* **2016**, *7*, 1.
[56] P. Jash, R. K. Parashar, C. Fontanesi, P. C. Mondal, *Adv. Funct. Mater.* **2022**, *32*, 2109956.
[57] P. Sachan, P. C. Mondal, *J. Mater. Chem. C* **2022**, *10*, 14532.
[58] R. Gupta, J. Pradhan, A. Haldar, C. Murapaka, P. Chandra Mondal, *Angew. Chemie Int. Ed.* **2023**, *62*, DOI 10.1002/anie.202307458.
[59] L. Chkoda, C. Heske, M. Sokolowski, E. Umbach, F. Steuber, J. Staudigel, M. Stößel, J. Simmerer, *Synth. Met.* **2000**, *111–112*, 315.
[60] P.-E. Hellberg, S.-L. Zhang, C. S. Petersson, *IEEE Electron Device Lett.* **1997**, *18*, 456.
[61] A. Novikov, *Solid. State. Electron.* **2010**, *54*, 8.
[62] F. Zahid, M. Paulsson, S. Datta, in *Adv. Semicond. Org. Nano-Techniques*, Elsevier, **2003**, pp. 1–41.

*Table of content*

**High Rectification Ratio at Room Temperature in Rhenium(I) Compound**

A Re(I) compound composed of π-stacking organic motifs utilized as electronic circuit elements in mimicking electrical current rectification with a three order of magnitude that operates at low-bias and at room temperature. The device is able to transform AC to DC signals at high frequency at 1 KHz, such electrical functions are highly desirable for practical applications.

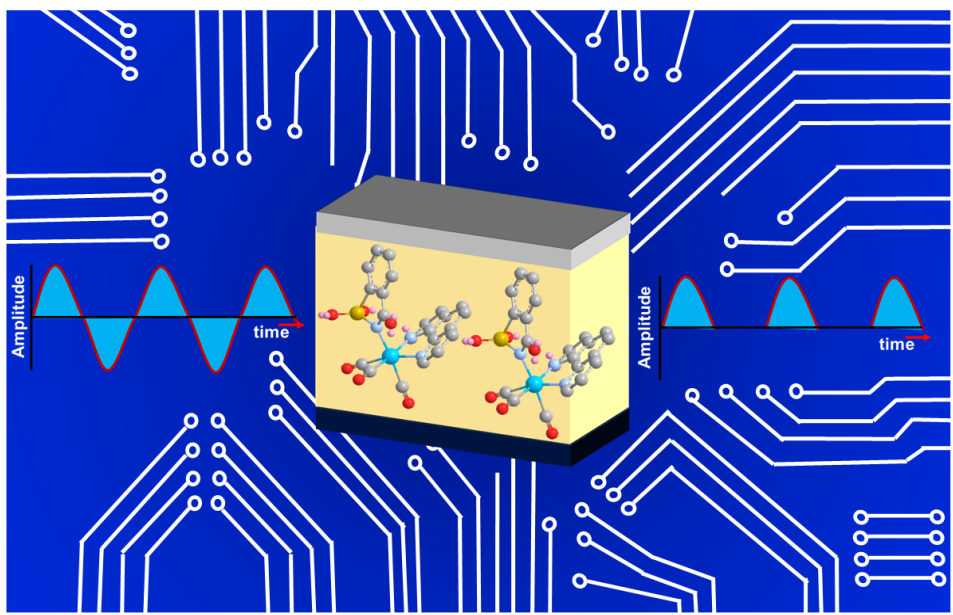